\documentclass[twocolumn]{revtex4}
\usepackage{graphicx}
\begin{document}

\title{Quantifying the resource of sharing a reference frame}
\author{S.J. van Enk$^{b,c}$}
\affiliation{$^b$Bell Labs, Lucent Technologies\\
600 Mountain Ave,\\ Murray Hill, NJ 07974
\\$^c$Institute for Quantum Information\\California Institute of Technology\\
Pasadena, CA 91125
}
\date{\today}

\begin{abstract}
We define a new quantity called {\em refbit}, which
allows one to quantify the resource of sharing a reference frame in quantum
communication protocols. By considering 
both asymptotic and nonasymptotic protocols
we find relations between {\em refbits}
and other communication resources.
We also consider 
the same resources in encoded, reference-frame independent, form.
This allows one to rephrase and 
unify previous work on phase references, reference frames,
and superselection rules.
\end{abstract}
\maketitle 
\section{Introduction}
The roles that reference frames play in 
communication protocols have attracted a lot of attention in recent times.
Many aspects related to the 
transmission of a direction in space by quantum particles
have been analyzed in detail, see for instance Refs.~\cite{massar}.
It was also pointed out that both classical and quantum communication are 
possible without sharing a reference frame by encoding information in
particular invariant subspaces \cite{bartlett2}.
Moreover, several papers have discussed
the relation between reference frames (or more generally, phase references) 
and super-selection rules,
and their role in quantum communication \cite{supers,super}.
Finally, sharing a secret reference frame as a cryptographic 
resource was analyzed in \cite{bartlett1}.

In most of the work mentioned above
a reference frame is assumed to be either fully present 
or fully absent. The present paper attempts to quantify the partial
presence of a phase reference.
Ref.~\cite{supers}, too, quantifies a resource that can substitute for
a phase reference. That quantity applies
to one case of two considered in the present paper (see Section \ref{model}),
and describes in fact two different resources,
the {\em ebit} and the {\em refbit}, which are similar in
certain contexts but different in others. 
The {\em ebit} and other resources, such as 
a unit of coherent communication, a {\em cobit}, were defined and analyzed
in Ref.~\cite{harrow}, but under the implicit assumption that the
communicating parties share a
reference frame.
Here we modify those definitions to explicitly take into account the absence
of a shared reference frame, and in addition
we
introduce
a unit of sharing a phase reference, a {\em refbit}.
The formalism presented here is an alternative, and hopefully useful,
 way of formulating
the role reference frames play in communication protocols.
Indeed, by following the methods of
\cite{harrow} we can rephrase and unify results from  several previous papers
on reference frames and superselection rules, such as
\cite{supers} and \cite{vaccaro} (see also \cite{rve}).
For example,
we will introduce encoded, phase-reference independent,
 versions of the resources,
such as an {\em Ebit}, which we will always denote by capitalizing the
word used for the unencoded resource.
The relations between {\em ebits} and {\em Ebits} 
clarify the various measures of entanglement used in \cite{vaccaro},
and also how quantum data hiding \cite{supers} in the presence of
superselection rules works, thus unifying those two results. 
The encoding used is inspired by that of Ref.~\cite{bartlett2}, 
but is different as our communication model, presented below, 
is different.

Since notation can be confusing as various terms, such as qubits and ebits,
are used in different contexts with different meanings, we start
out by clarifying the notation used in this paper.
\section{Notation}\label{not}
A two-level atom or a polarized photon can act as a physical qubit.
On the other hand, the {\em qubit} used in this paper (and
in Ref.~\cite{harrow}) is a communication resource, and is equivalent
to sending a physical qubit over a noiseless channel (this will be made
more precise in Section \ref{def}). In order to distinguish
the two types of ``qubits'', 
we will always write the communication resource {\em qubit} in italic,
the other type of ``qubit'' will always be prefaced by the 
word ``physical'' and written
in roman.

Somewhat similarly, the ebit is known as a unit of entanglement.
It can be used to quantify the amount of entanglement in any 
bipartite state. The {\em ebit} used in this paper (and also in \cite{harrow})
is, again, a communication resource. For example,
when Alice and Bob are engaged
in some quantum communication protocol, then an {\em ebit}
is the resource of Alice and Bob sharing an entangled state of a 
particular form. That entangled state contains one ebit of entanglement.

Finally, the following notation should be clear by now:
a classical bit is a unit of classical information,
the {\em cbit} in the present paper is a communication resource
and corresponds to
sending a classical bit over a noiseless channel.
\section{Communication model}\label{model}
Assume that Alice and Bob agree on the definition of the 
qubit states $|0\rangle$ and $|1\rangle$, but not
on the definition of the phase $\phi$ in superpositions of the form
\[ \sin\alpha |0\rangle +\cos\alpha e^{i\phi}|1\rangle.\]
For them to agree on the value of $\phi$ they would have to share 
a phase reference. 

This model corresponds to Alice and Bob communicating with either
photon number states, with $|0\rangle$ and $|1\rangle$ denoting
states containing no and one photon in one particular mode \footnote{
The assumption here is, of course, that particular modes can be defined 
unambiguously without
sharing a reference frame: for that purpose, too, we can use circular 
polarization, or the modes with cylindricially symmetric transverse 
mode profiles discussed in \cite{fed}.}, or
with polarized single photons, with $|0\rangle$ and $|1\rangle$
denoting left-hand and right-hand circular polarization.
The phase references needed in these two cases are a synchronized clock
and a spatial reference frame, respectively \footnote{
In both implementations Alice and Bob do need to share something, namely 
knowledge of what they mean by ``0'' and ``1'', or by ``left'' and ``right''.
That resource, though, seems sufficiently mild to allow it without further
consideration}.

Both Alice and Bob are assumed
to have local phase references 
at their disposal, which define local phases $\phi_A$ and $\phi_B$, 
respectively. 
It is important to note that these two phases in turn are defined
only with respect to another, fictituous phase reference, that we
may assume to be in the hands of a third party.
For instance, 
in the context of
cryptographic protocols we may assume this third party is an eavesdropper.
Neither Alice nor Bob are aware of the values of $\phi_A$ and $\phi_B$.
Thus, the third party reference frame has a privileged role. 

Those who believe that superselection rules
might forbid coherent superpositions of $|0\rangle$ and $|1\rangle$ \footnote{
Presumably, the superselection rule 
would apply only in the context of using different number 
states, not in the case
of using different polarizations.}
may prefer the following, alternative, cumbersome, but in the end equivalent, 
formulation: 
The fictituous third party has a state of the form
\[ \int \frac{{\rm d}\phi}{2\pi}
((|0\rangle+\exp(i\phi)|1\rangle)
(\langle 0|+\exp(-i\phi)\langle 1|))^{\otimes N},\]
with $N\rightarrow\infty$. In spite of appearances
this state contains no coherent
superpositions of $|0\rangle$ and $|1\rangle$. 
Now whenever a phase between $|0\rangle$ and $|1\rangle$ appears in some 
equation 
this phase is understood to be relative to the dummy phase $\phi$. 
For example, the equivalent of
''Alice having a state
$|0\rangle_A+\exp(i\phi_A)|1\rangle_A$'' is then that the joint state 
of the third party and Alice's qubit is
\begin{eqnarray*} 
\int \frac{{\rm d}\phi}{2\pi}
((|0\rangle+\exp(i\phi)|1\rangle)
(\langle 0|+\exp(-i\phi)\langle 1|))^{\otimes N}\otimes\\
(|0\rangle+\exp(i(\phi+\phi_A))|1\rangle)
(\langle 0|+\exp(-i(\phi+\phi_A))\langle 1|).\end{eqnarray*}
One can define in a similar way what it means for Alice and Bob to have their 
own phase references:
large collections of physical
qubits with phases $\phi+\phi_A$ or $\phi+\phi_B$, 
respectively.

In any case,  
Alice and Bob can perform any local operation they like, except
that there will always be extra phase factors $\exp(i\phi_A)$
or $\exp(i\phi_B)$ appearing
in front of the states $|1\rangle_A$ and $|1\rangle_B$.
For example, consider the Hadamard transformation. 
When Alice performs her version of that transformation, 
 she actually performs
$H_A$:
\begin{eqnarray}
|0\rangle&\mapsto& |0\rangle +\exp(i\phi_A)|1\rangle\nonumber\\
|1\rangle&\mapsto& \exp(-i\phi_A)|0\rangle +|1\rangle,
\end{eqnarray}
as described from the third-party reference frame.
Similarly, when she performs her local version of a controlled-NOT, 
she actually
performs CNOT$_A$:
\begin{eqnarray}
|0\rangle|0\rangle&\mapsto& |0\rangle|0\rangle\nonumber\\
|0\rangle|1\rangle&\mapsto& |0\rangle|1\rangle\nonumber\\
|1\rangle|0\rangle&\mapsto& \exp(i\phi_A)|1\rangle|1\rangle\nonumber\\
|1\rangle|1\rangle&\mapsto& \exp(-i\phi_A)|1\rangle|0\rangle.
\end{eqnarray}
The three Pauli operations become
\begin{eqnarray}
Z_A&=&Z= \left(
\begin{array}{cc}
1&0\\ 0&-1 
\end{array}
\right)\nonumber\\
X_A&=& \left(
\begin{array}{cc}
0&\exp(i\phi_A)\\ \exp(-i\phi_A)&0 
\end{array}
\right)\nonumber\\
Y_A&=& \left(
\begin{array}{cc}
0&\exp(i\phi_A)\\ -\exp(-i\phi_A)&0 
\end{array}
\right)
\end{eqnarray}
If we picture a (Bloch) sphere representing a qubit, then
Alice and Bob agree on the north and south poles, they agree on the latitude
of all points on the sphere, but not on the longitude. 
They can perfectly define rotations around the polar axis, but not 
around any other rotation axis. 
\section{Definitions of resources}\label{def}
In order to define various resources such as {\em ebits}
 and {\em qubits}, we follow
Ref.~\cite{harrow}, but with the appropriate modifications to
reflect the assumptions of the communication model 
defined in the preceding Section.
Ref.~\cite{harrow} implicitly assumed Alice and 
Bob do share a phase reference. The first three definitions given below 
refer to resources that are {\em sent} from one party to the other 
(and in the definitions we always 
consider Alice the sender and Bob the receiver). 
The last three definitions
define resources {\em shared} by Alice and Bob.

Before we give the definitions it is perhaps useful to point out the 
following. Under the assumption of a shared reference frame 
the definition of an {\em ebit} (as in \cite{harrow}), 
for example, includes a more general
class of entangled states than the ones considered below. 
Moreover, entangled states are more 
powerful resources in the presence of a reference frame.
Indeed, that was one point of Refs.~\cite{supers} and \cite{vaccaro}.
The {\em ebit} as defined in the present paper is less
powerful, but the {\em Ebit} 
defined later on is  much more similar
to the quantity defined in \cite{harrow}.

\begin{enumerate}
\item
The most powerful resource is a {\em qubit}, that is, one use of a perfect 
quantum communication channel. It is the ability to send
any physical qubit from Alice to Bob, i.e.,
\[ 
a|0\rangle_A+b|1\rangle_A\mapsto 
a|0\rangle_B+b|1\rangle_B. 
\]
Here the coefficient $b$ does not necessarily contain an explicit phase factor 
$\exp(i\phi_A$) as the physical
qubit may have been handed to Alice by a third party 
(this may be relevant in a teleportation protocol, for instance).
\item
A {\em cobit} is the ability to perform
\begin{eqnarray*}
|0\rangle_A&\mapsto& 
|0\rangle_A |0\rangle_B,\\ 
|1\rangle_A&\mapsto& 
\exp(i\phi_A)|1\rangle_A |1\rangle_B. 
\end{eqnarray*}
Here the phase factor $\exp(i\phi_A)$ does always appear.
This definition is basis-dependent, as it singles out the 
$|0\rangle,|1\rangle$ basis. This basis dependence
is necessary as the {\em cobit}
is defined by a cloning-like operation.
\item A {\em cbit}, corresponds to one
use of a classical communication channel, which can be described by the 
process
\begin{eqnarray*}
|0\rangle_A&\mapsto& 
|0\rangle_E |0\rangle_B,\\ 
|1\rangle_A&\mapsto& 
\exp(i\phi_A)|1\rangle_E |1\rangle_B, 
\end{eqnarray*}
where $E$ refers to the environment, assumed
unobservable by either Alice or Bob, but possibly observable
by an eavesdropper. In any protocol involving {\em cbits},
we thus assume that Alice and Bob trace out
the environment.
\item The following definition of an {\em ebit}
is chosen such that
both a single {\em qubit} and a single {\em cobit} can be used to generate 
an {\em ebit}, a bipartite entangled state
of a particular form, shared by Alice and Bob:
Alice, starting out with the state 
$|0\rangle\pm\exp(i\phi_A)|1\rangle$, can use a {\em cobit} to produce
\[
|0\rangle_A|0\rangle_B\pm\exp(2i\phi_A)|1\rangle_A|1\rangle_B.
\]
By means of a local operation (a bit flip $X_A$)
Alice can convert this state to
\[
|1\rangle_A|0\rangle_B\pm|0\rangle_A|1\rangle_B,
\]
apart from an irrelevant overall phase factor.
The latter state can also directly be produced by either Alice or Bob
by creating that state locally and subsequently using a {\em qubit}.
Starting from an {\em ebit} as above, Bob, too, can apply local operations 
and produce
\[
\exp(2i\phi_A)|1\rangle_A|0\rangle_B\pm\exp(2i\phi_B)|0\rangle_A|1\rangle_B.
\]
Since all these entangled states are connected by local operations, these 
are all equivalent definitions of an {\em ebit}.
\item
A {\em refbit} is, like an {\em ebit}, 
a shared resource between Alice and Bob. 
It is
 defined as  Alice and Bob sharing a (product) state of the form
\[ 
(|0\rangle_A+\exp(i\phi_A)|1\rangle_A)(
|0\rangle_B+\exp(i\phi_A)|1\rangle_B),
\]
or equivalently
\[ 
(|0\rangle_A+\exp(i\phi_B)|1\rangle_A)(
|0\rangle_B+\exp(i\phi_B)|1\rangle_B).
\]
This definition is such that
Alice can use a {\em qubit} to establish a {\em refbit}.
Since neither Alice nor Bob are aware of any of the phases appearing 
in the definition of a {\em refbit}, 
they cannot establish a {\em refbit} by 
local means, in spite of it being a product state.
\item For later use, we also define a 
{\em refbit(2)}.
Again, this is a shared resource between Alice and Bob. 
It is
 defined as  Alice and Bob sharing a state of the form
\[ 
(|00\rangle_A+\exp(2i\phi_A)|11\rangle_A)(
|00\rangle_B+\exp(2i\phi_A)|11\rangle_B).
\]
One can probabilistically generate 
a {\em refbit(2)} from 2 {\em refbits} (succeeding with 50\% chance), but
not the other way around.
\end{enumerate}
In addition to the above resources, which do not use any coding,
we can also define encoded versions of the same resources.
In particular, Alice and Bob can communicate without sharing a reference frame
using the encodings discussed in \cite{bartlett2}.
That paper actually discusses a slightly different situation where 
Alice and Bob do not even agree on the definitions
of $|0\rangle$ and $|1\rangle$.  In that case,
which would correspond to the scenario
of Alice and Bob
using {\em massive} spin-1/2 particles to communicate, 
one needs 4 physical qubits 
to encode one logical qubit.
On the other hand, within the present communication model, 
Alice and Bob can
encode in the following reference-frame independent way, using just
two physical qubits
to encode one logical qubit. For example, they could use
\begin{eqnarray}
|0\rangle_L&=&|0\rangle|1\rangle,\nonumber\\
|1\rangle_L&=&|1\rangle|0\rangle.
\end{eqnarray}
Alice can encode a logical qubit by performing
\begin{eqnarray}
|0\rangle(a|0\rangle+b\exp(i\phi_A)|1\rangle)
\mapsto
\exp(i\phi_A)
(a|0\rangle_L+b|1\rangle_L).
\end{eqnarray}
We denote resources that make use of this type of encoding by
capitalizing the corresponding unit. 
So we have {\em Qubits}, {\em Cbits}, {\em Ebits}, and 
{\em Cobits}, but
{\em Refbits} do not make any sense. 
\section{Relations between resources}
With these definitions  we can write down simple relations
of the form $X\geq Y$, following \cite{harrow}, which mean that the resource 
$X$ can be used to simulate resource $Y$ (and $X=Y$
if and only if both $X\geq Y$ and $Y\geq X$).

We consider first coherent protocols, in which the environment, 
which appears in the definition of a {\em cbit}, plays no role. Subsequently,
we consider 
incoherent protocols that (implicitly or explicitly) yield or use {\em cbits}.
\subsection{Coherent protocols}
With the above definitions it is straightforward 
to obtain protocols that achieve 
\begin{eqnarray}
1\, qubit&\geq& 1\, ebit,\nonumber\\
1\, qubit&\geq& 1\, cobit,\nonumber\\
1\, cobit&\geq& 1\,  ebit,\nonumber\\
1\, qubit&\geq& 1\,  refbit,\nonumber\\
1\, qubit+1\,ebit&\geq& 1\, refbit(2).
\end{eqnarray}
We now consider several slightly more complicated
protocols that convert one type of resource into another.
\subsubsection{Protocol C1}
First, consider a protocol that is simpler than teleportation,
yet achieves the same resource-wise.
Alice is given a 
physical
qubit in the state (possibly unknown to her) 
$a|0\rangle_A+b|1\rangle_A$. 
Using a {\em cobit}, she accomplishes
 \[a|0\rangle_A+b|1\rangle_A
\mapsto
a|0\rangle_A|0\rangle_B+
b\exp(i\phi_A)|1\rangle_A|1\rangle_B.
\]
Then on her half of this state plus an ancilla
in state $|0\rangle$ she performs the following 2-qubit operation
\begin{eqnarray*}
|0\rangle |+\rangle &\mapsto& |0\rangle |+\rangle,\\
|0\rangle |-\rangle &\mapsto& \exp(i\phi_A)|1\rangle |-\rangle,
\end{eqnarray*}
where $|\pm\rangle=|0\rangle\pm \exp(i\phi_A)|1\rangle$.
This leads to the state
\begin{eqnarray*}
|0\rangle_{A1} |+\rangle_A   (a|0\rangle_B+b|1\rangle_B)+\\
\exp(i\phi_A)|1\rangle_{A1} |-\rangle_A   (a|0\rangle_B-b|1\rangle_B).
\end{eqnarray*}
Subsequently, Alice uses another {\em cobit} to copy her ancilla $A1$ 
into an ancilla $B1$ at Bob's site. He then performs the operation
\begin{eqnarray*}
|0\rangle |1\rangle &\mapsto& |0\rangle |1\rangle,\\
|1\rangle |1\rangle &\mapsto& -|1\rangle |1\rangle,
\end{eqnarray*}
to end up with the state
\begin{eqnarray*}
(|0\rangle_{A1}|+\rangle_A |0\rangle_{B1}
+\exp(i\phi_A) |1\rangle_{A1} |-\rangle_A |1\rangle_{B1})\\
\otimes(a|0\rangle_B+b|1\rangle_B).
\end{eqnarray*}
The first term can be converted into one {\em ebit} by Alice, 
the second term shows 
Alice has managed to simulate a {\em qubit}. Thus, this protocol achieves
\begin{equation}
2\, cobits
\geq 1\, qubit+1\,  ebit.
\end{equation}
First we note Alice and Bob do not have to share a reference frame or 
any {\em refbits} to achieve this.
Second we note this relation holds without using a catalyst,
thus slightly improving Eq.~(4) in \cite{harrow},
which was derived there using an extra
{\em ebit} on both sides of the relation. 
\subsubsection{Protocol C2}
A similar protocol starts with Alice producing
\[a|0\rangle_A+b\exp(i\phi_A)|1\rangle_A.\]
Using a {\em cobit} Alice and Bob share the state
\[a|0\rangle_A|0\rangle_B+b\exp(2i\phi_A)|1\rangle_A|1\rangle_B.\]
Then Alice flips her physical qubit to get
\[a|1\rangle_A|0\rangle_B+b|0\rangle_A|1\rangle_B.\]
apart from an irrelevant overall phase factor $\exp(i\phi_A)$.
Sending her physical qubit to Bob (and thus using one {\em qubit}) 
leaves him 
with an encoded {\em Qubit}. Thus, we find
\begin{equation}\label{Qubit}
1\, cobit +1\, qubit 
\geq 1\,Qubit.
\end{equation}
\subsubsection{Protocol C3}
By starting out with a {\em refbit}, $|0\rangle+\exp(i\phi_B)|1\rangle$, Alice
can use a {\em cobit} to produce
\[|0\rangle_A|0\rangle_B+\exp(i\phi_A+i\phi_B)|1\rangle_A|1\rangle_B,\]
which by local transformations can be transformed into an {\em Ebit}. Hence
\begin{equation}
1\, cobit+
 1\, refbit
\geq 1\, Ebit.
\end{equation}
Note that starting with just one {\em ebit}, Alice and Bob can 
generate by local operations the state
\[
\exp(i\phi_A)|0\rangle_L|1\rangle_L
+\exp(i\phi_B)|1\rangle_L|0\rangle_L\]
but that state contains both $\phi_A$ and $\phi_B$, and so is not a 
reference-frame invariant state, and hence not an {\em Ebit}.
\subsubsection{Protocol C4}
Instead of using  a {\em refbit} and a {\em cobit} to obtain an 
{\em Ebit}, 
it is easy to check that Alice and Bob could also
start out with an {\em ebit} and then use a {\em cobit}
 to end up with an {\em Ebit}, thus 
leading to
\begin{equation}\label{Ebit}
1\, cobit+
 1\, ebit
\geq 1\, Ebit.
\end{equation}
\subsubsection{Superdense coding}
Finally, consider the coherent version of superdense coding.
Alice and Bob start with an {\em ebit}, say 
$|1\rangle_A|0\rangle_B+|0\rangle_A|1\rangle_B$.
Alice, moreover, has two ancilla physical
qubits in a state $|a_1\rangle|a_2\rangle$,
where $a_1$ and $a_2$ take on the values 0 or 1.
She then performs  the following 3-qubit operation, conditioned
on the state of the two ancilla's (the first two kets in the equations below)
\begin{eqnarray*}
|1\rangle|0\rangle|0\rangle&\mapsto&|1\rangle|0\rangle|0\rangle\nonumber\\
|1\rangle|0\rangle|1\rangle&\mapsto&|1\rangle|0\rangle|1\rangle\nonumber\\
|0\rangle|1\rangle|0\rangle&\mapsto&|0\rangle|1\rangle|0\rangle\nonumber\\
|0\rangle|1\rangle|1\rangle&\mapsto&-|0\rangle|1\rangle|1\rangle\nonumber\\
|0\rangle|0\rangle|0\rangle&\mapsto&
\exp(i\phi_A)|0\rangle|0\rangle|1\rangle\nonumber\\
|0\rangle|0\rangle|1\rangle&\mapsto&
\exp(-i\phi_A)|0\rangle|0\rangle|0\rangle\nonumber\\
|1\rangle|1\rangle|0\rangle&\mapsto&-
\exp(i\phi_A)|1\rangle|1\rangle|1\rangle\nonumber\\
|1\rangle|1\rangle|1\rangle&\mapsto&
\exp(-i\phi_A)|1\rangle|1\rangle|0\rangle\nonumber\\
\end{eqnarray*}
Subsequently, Alice uses one {\em qubit} to send
the physical qubit $A$ to Bob.
In the four possible cases Alice and Bob
end up with 
\[|1\rangle|0\rangle\otimes (
|1\rangle|0\rangle+|0\rangle|1\rangle)\]
\[|0\rangle|1\rangle\otimes (
-|1\rangle|0\rangle+|0\rangle|1\rangle)\]
\[|0\rangle|0\rangle\otimes (
\exp(-i\phi_A)|0\rangle|0\rangle+\exp(i\phi_A)|1\rangle|1\rangle)\]
\[|1\rangle|1\rangle\otimes (
\exp(-i\phi_A)|0\rangle|0\rangle-\exp(i\phi_A)|1\rangle|1\rangle)\]
The last two entangled states are indistinguishable to Bob. 
More precisely, both are equal mixtures of $|0\rangle|0\rangle$
and $|1\rangle|1\rangle$ to him. One of the two classical
bits $a_1$ and $a_2$, 
therefore, cannot be sent coherently
without a reference frame, but the other bit can.
Thus, by starting off her ancilla bits in only two possible states,
corresponding to $|0\rangle_L$ or $|1\rangle_L$, Alice and Bob achieve
\begin{equation}
1\,qubit+1\, ebit\geq 
1\, Cobit. 
\end{equation}
\subsubsection{Comparing encoded and unencoded resources}
Performing superdense coding and protocol C1 with encoded resources
immediately gives us
\begin{equation}
1\,Qubit+1\, Ebit=
2\, Cobits. 
\end{equation}
These encoded resources can be obtained from unencoded resources,
by the results obtained above. In particular, the left-hand
side, 
a {\em Qubit} and an {\em Ebit}, 
can be obtained from 2 {\em cobits}, 1 {\em qubit}, and 1 {\em ebit}.
The right-hand side can indeed be obtained from the same
unencoded resources. Namely,
the two {\em cobits} can be converted into 1 {\em qubit}
 and 1 {\em ebit}. Subsequently,
2 {\em qubits} and 2 {\em ebits} can be converted into 2
{\em Cobits}.

Obviously, we can also use 2 {\em cobits} directly to yield one 
{\em Cobit}.
From this and Eqs~(\ref{Qubit}) and (\ref{Ebit}) 
one sees that a {\em cobit} can always be used
to convert an unencoded resource, a {\em qubit}, an {\em ebit},
and a {\em cobit}, into the corresponding encoded form.

We also note that unencoded resources cannot be obtained from just encoded
resources. The reason is simply that unencoded resources
contain formation about Alice's and Bob's local reference frames, while
encoded resources do not (which is the whole point of encoding).

\subsection{Incoherent protocols}
\subsubsection{Superdense coding}
Returning to superdense coding, we note that Alice 
could decide to start off her ancilla bits in only three possible initial
states, 01, 10 or 00, with equal probabilities. 
In the incoherent version she thus succeeds in
sending $\log_2(3)$ classical bits to Bob, rather than 2 when they share 
a reference frame. 
Moreover, in the case that Alice chose 00, Bob actually ends up
with a state in his possession that still contains the phase $\phi_A$, 
$|00\rangle+\exp(2i\phi_A)|11\rangle$.
Clearly, this can be 
converted into a {\em refbit(2)}. Thus, Alice and Bob actually achieve
\begin{equation}\label{23}
1\,qubit+1\, ebit\geq 
\log_2(3)\, cbits+1/3\,  refbit(2). 
\end{equation}
Now, with one {\em refbit} Bob cannot get more {\em
cbits} out of this protocol,
basically since he has to
compensate for extra phases of $2\phi_A$ appearing. But
with two {\em refbits} he can do a better job of 
decoding in the classical case, and, moreover, can sometimes save his
{\em refbits}. In particular, Bob has a probability $P=1/4$ to
unambiguously discriminate between the two states
\[
(|00\rangle\pm\exp(2i\phi_A)|11\rangle)\otimes 
(|0\rangle+\exp(i\phi_A)|1\rangle)^{\otimes 2}.\]
(There are two ways of obtaining the probability $P=1/4$:
either we write down mixtures over the unknown phase $\phi_A$ for the 
two possible states
and calculate the unambiguous-state discrimination probability for 
two mixed states, or we let Bob project onto subspaces
with equal phase factors $\exp(in\phi_A)$, where $n=0,1,2,3,4$,
and subsequently let him perform unambiguous state 
discrimination between the projections of the two states 
within those subspaces.)
Hence, 
if Alice chooses her two classical bits as either $01$ or 
$10$ with probability $p/2$ each, and as $00$ or $11$ with probability 
$(1-p)/2$ each, then they achieve
\begin{eqnarray}
1\, qubit+1\,  ebit +2\, refbits\geq \nonumber\\ 
(H(p)+(1+3p)/4)\,  cbits +2p\,  refbits,
\end{eqnarray}
with $H(p)$ the Shannon entropy $H(p)=-p\log_2p-(1-p)\log_2(1-p)$.
Note that in this case no further resources are left over, 
in particular, there is 
no {\em refbit(2)} at the end of the protocol.
Choosing $p=2/3$ we gain 1/12 of a {\em cbit} compared to (\ref{23}) 
while using up 2/3 {\em refbits}:
\begin{eqnarray}
1\,qubit+1\,  ebit +2\, refbits\geq \nonumber\\
(\log_23+1/12)\,  cbits +4/3\,  refbits.
\end{eqnarray}
The amount of communicated {\em cbits}
 using two {\em refbits} is maximized to
1.6732 {\em cbits} achieved for $p=p_0\approx 0.627$.

Instead of using two {\em refbits}, 
it is better for Bob to use one {\em refbit(2)}.
It is easy to verify that Bob now can succeed in unambiguously
distinguishing the two states
\[
(|00\rangle\pm\exp(2i\phi_A)|11\rangle)\otimes 
(|00\rangle+\exp(2i\phi_A))|11\rangle\]
with probability $P=1/2$. This then yields
\begin{eqnarray}
1\, qubit+1\,  ebit +1\, refbit(2)\geq \nonumber\\
(H(p)+(p+1)/2)\, cbits +p\,  refbits(2).
\end{eqnarray}
We can also calculate what Alice and Bob can gain from sharing many
{\em refbits}. The important determining factor here is the probability
$P_N$ for Bob to unambiguously discriminate between the states
\[
(|00\rangle\pm\exp(2i\phi_A)|11\rangle)\otimes 
(|0\rangle+\exp(i\phi_A)|1\rangle)^{\otimes N},\]
when Alice and Bob share $N$ refbits.
It is straightforward to find for even values of $N$
\begin{equation}
1-P_N=\frac{
\left(
\begin{array}{c}
N\\N/2 
\end{array}
\right)+
\left(
\begin{array}{c}
N\\N/2+1 
\end{array}
\right)
}{2^N}
\end{equation}
while for odd values of $N$ one finds the same result as for $N-1$. That is,
a single extra {\em refbit} never helps to improve upon the case with an even
number of {\em refbits}.
So we find that $P_N$ approaches unity only slowly. Asymptotically one has
\[P_N\approx 1-\frac{4}{\sqrt{2\pi N}}.\]
In terms of $P_N$ and $p$, superdense coding leads to
the following trade-off relation
\begin{eqnarray}\label{super}
1\, qubit+1\,  ebit +N\, refbits\geq \nonumber\\ 
(H(p)+(1-P_{N})p+P_{N})\,  cbits +pN\,  refbits.
\end{eqnarray}
As expected, in the limit of $N\rightarrow\infty$ one recovers standard
superdense coding in the presence of a shared reference frame:
in particular, one uses $p=1/2$
to send 2 {\em cbits},
and one saves half of the {\em refbits}.
\subsubsection{Teleportation}
The coherent version of teleporation works only if Alice and Bob 
share a reference frame or use phase-invariant encoding.
However, it is still interesting to consider the resources needed
for the incoherent (standard)
version of teleportation.
Without a reference frame teleportation only succeeds for two out of four
outcomes of Alice's Bell measurement, thus leading to
\begin{equation}\label{tel}
2\,  cbits
+1\,  ebit\geq 
1/2\, qubit. 
\end{equation}
Here we are interested only in perfect fidelity teleportation, 
and in only half of the cases
do they succeed in this endeavor
(and they know when they succeed and when not).
When Alice and Bob share {\em refbits}, the probability to succeed in 
perfect teleportation increases. Namely, in the cases where Alice
gets a ``wrong'' measurement outcome, Bob can still
try to project
onto the correct subspace. The probability to succeed turns out to
be the same probability $P_N$ as we encountered before
when we considered superdense coding. Thus we find the duality
between the two protocols persists in the absence of a reference frame
and sharing an arbitrary number of {\em refbits}. Here we get (for even numbers
of {\em refbits})
\begin{equation}\label{telr}
2\, cbits
+1\, ebit+2N\,  refbits\geq 
(P_{2N}+1)/2\, qubit+N\, refbits. 
\end{equation}
For example, 
with 2 {\em refbits} 
one succeeds in perfect teleportation with probability
5/8, but, as before,
when using a {\em refbit(2)}, the chances increase, and one gets
\begin{equation}\label{tel2}
2\,  cbits
+1\, ebit+1\,  refbit(2)\geq 
3/4\, qubit+1/2\, refbit(2). 
\end{equation}
In the limit $N\rightarrow\infty$ one recovers
the results for teleportation in the presence of a shared 
reference frame, and again one saves half of the {\em refbits}.
\subsubsection{Converting ebits to Ebits}
In this subsection we consider some more protocols
that convert unencoded into encoded resources.
These protocols in fact use the present formalism to
reformulate
 and unify results obtained before in Refs.~\cite{supers} and \cite{vaccaro}.
Starting out with 2 {\em ebits}, 
$(|0\rangle|1\rangle+|1\rangle|0\rangle)^{\otimes 2}$,
Bob can
perform a projective measurement onto subspaces with
even or odd numbers of $|1\rangle$ appearing in his two-qubit space.
This leads with probability 1/2 to an {\em Ebit}, and with probability 1/2
to a state that one would normally called entangled, but which is not
equivalent to an {\em Ebit}. Thus, we get
\begin{equation}
2\, ebits\geq 
1/2\,  Ebit. 
\end{equation}
This relation in fact reexpresses the statement of Eq.~(5) in 
the second paper of Ref.~\cite{vaccaro}, namely, that
the amount of entanglement $E_P$, for 2 {\em ebits} is 1/2, although it is zero
for 1 {\em ebit}.
We can achieve the same conversion with a {\em refbit}: 
\begin{equation}
1\, ebit+1\, refbit
\geq 1/2\,  Ebit. 
\end{equation}
Thus for the purpose of extracting ''useful'' entanglement out of 
an {\em ebit},
another {\em ebit} or a {\em refbit} achieve the same.
Both relations also express how data hiding works in the presence
of superselection rules. Namely, an {\em ebit} can be used to hide one 
classical bit, by
encoding it in one of two states $|0\rangle_A|1\rangle_B\pm 
|1\rangle_A|0\rangle_B$.
However, neither Alice nor Bob can locally distinguish these two states 
at all. But by using either a {\em refbit} or a second {\em ebit}, they can 
convert it
with 50\% probability to one of two different forms 
of an {\em Ebit}. Since the latter
is in encoded form, the classical bit can be retrieved.

We might as well remark here that although {\em refbits} and 
{\em ebits} are similar 
for certain purposes (see, for example, \cite{supers}), 
they are definitely not the same. In particular,
a {\em refbit} cannot be converted into an {\em ebit} by local operations and
classical communication, but 
one can achieve
the converse by remote state preparation,
\begin{equation}
1\, ebit+1\, cbit
\geq 1\, refbit. 
\end{equation}
Also, whereas an {\em ebit} can be converted into a secret shared random 
classical bit,
a {\em refbit} cannot.
  
 Alice and Bob
 can reach the optimum 
conversion of 1 {\em ebit} to 1 {\em Ebit}
in the limit of infinitely many
{\em refbits}. Bob projects onto subspaces with a 
fixed number of states $|1\rangle$ on his $(N+1)$ physical qubits 
(one for the {\em ebit}, the others from the $N$ {\em refbits}
 he shares with Alice).
The outcomes do not always lead to maximally entangled encoded states, 
but they do always lead to encoded states. The average entanglement they gain
with 2 {\em refbits}, for example, is
\begin{equation}
1\, ebit+2\, refbits
\geq (3\log_2(3)/4-1/2)\, Ebit. 
\end{equation}
By using a large number $N$ of {\em refbits}, 
they approach a full {\em Ebit},
according to
\begin{equation}
1\, ebit+N\, refbits
\geq (1-1/(2N))\,  Ebit. 
\end{equation}
\subsection{Asymptotic relations}
So far we only considered single-shot protocols, in which a single use 
of certain resources is analyzed, but 
one may also be interested in the results that can be obtained
with many uses of a given resource. For example,  let us consider
incoherent superdense coding.
Without {\em refbits}, Alice and Bob in a single-shot protocol can achieve
\begin{equation}
1\, qubit+1\, ebit\geq 
\log_2(3)\,  cbits+1/3\,  refbit(2). 
\end{equation}
They achieve this if Alice applies one of three operations to her 
physical qubit,
$I,Z,X_A$.
But if they use 2 {\em qubits} and 2 {\em ebits} the same protocol can yield
 slightly more than
twice the right-hand side. For instance, suppose Alice still
applies one of the same three operations $I,Z,X_A$ on the first 
physical qubit, 
with probability 1/3 each.
On her second physical qubit, 
she also applies $I$ or $Z$ with probability 1/3 each,
but she applies now either $X_A$ or $Y_A$ with probability 1/6 each.
Now the combination $X_A^{(1)}X_A^{(2)}$
can be unambiguously distinguished by Bob from
$X_A^{(1)}Y_A^{(2)}$ with probability P=1/2.
Thus, they gain 1/18th of a classical bit, 
but end up with only 2/9th of a {\em refbit(2)}.
Thus 
\begin{equation}
2\, qubits+2\,  ebits\geq 
(\log_2(3)+\frac{1}{18})\,  cbits+2/9\,  refbit(2). 
\end{equation}
The protocol above is just to illustrate that Alice and Bob can gain classical
bits,
it is not optimized for anything in particular. It does illustrate
how precious Hilbert space is wasted. It is much better indeed to extend the 
phase-reference-invariant encoding used above to higher dimensions.
It is trivial to accomplish this: for a fixed number $N$ of physical
qubits
we choose states with some fixed number $N_1$ of 1s, where
$N_1\approx [N/2]$. The dimension of the
subspace spanned by states of that form is $N!/(N1!(N-N1)!)$.
For large $N$ this number approaches 
\[
\frac{N!}{N1!(N-N1)!}\approx \frac{2^{N}}{\sqrt{\pi N/2}}.
\]
That means we can encode $N-\log_2(\pi N/2)$ encoded {\em Qubits} into
$N$ {\em qubits}. Asymptotically, therefore, we will have
\begin{equation}
1\, qubit\geq 
1\, Qubit\,\,\,(a), 
\end{equation}
with $(a)$ denoting the relation holds asymptotically for many copies
of the resources.
Similarly, suppose Alice and Bob start out with a large number $N$ of 
{\em ebits}. 
Alice and Bob can do a projective measurement on subspaces spanned by
states with a fixed number of 1s. Both will find with high probability a 
subspace with about $[N/2]$ 1s. Thus, they end up with high probability
(approaching unity) $N-\log_2(\pi N/2)$ {\em Ebits}. Thus,
\begin{equation}
1\, ebit\geq 
1\, Ebit\,\,\,(a). 
\end{equation}
This relation in fact reexpresses a fact analyzed in \cite{supers}.
As a result, superdense coding leads to
\begin{equation}
1\, qubit+1\,  ebit\geq 
2\, cbits\,\,\,(a). 
\end{equation}
\section{Summary and Discussion}
We presented a formalism for describing resources in quantum communication
that extends that of Ref.~\cite{harrow} to allow for the absence of
shared reference frames. It naturally leads to the definition of a 
new resource, the {\em refbit}, and to modifications of the 
definitions of known resources, such as the {\em ebit} and the 
{\em cobit}.

The formalism can be used to describe both known theoretical results and
practical experiments. For example, the relation
\begin{equation}
2\, ebits\geq 
1/2\, Ebit
\end{equation}
has the same meaning as the statement in \cite{vaccaro} that the
entanglement $E_P$ in two copies of the state
$|0\rangle|1\rangle+|1\rangle|0\rangle$ is 1/2. That is, in the presence
of a $U(1)$-superselection rule one needs two copies of
that state in order to generate {\em accessible} entanglement
with probability 1/2.
It in fact also expresses the fact that quantum data hiding  
in the presence of superselection rules \cite{supers} 
allows one to hide one classical bit of information
in an {\em ebit}, which can be unlocked with probability 1/2 if
another {\em ebit} is used as a resource.

To give a practical example described by our formalism,
consider today's quantum key distribution protocols.
Probably the best method to 
encode quantum information is in the relative phase of weak coherent states.
That is, in the standard implementation of the BB84 protocol
Alice sends Bob one of four states
\begin{equation}\label{qkd}
|\alpha \exp(i\phi_A)\rangle
|\alpha \exp(i(\phi_A+\phi_k))\rangle\,\,k=1\ldots4,
\end{equation}
where $\phi_k=k\pi/2$. Here $|\alpha\rangle$ denotes a 
coherent state with amplitude $\alpha$, where 
typically $|\alpha|^2\approx 0.1$.
Even if polarization is used to encode information in weak coherent states, 
one can still rewrite the states used in the form (\ref{qkd}) \cite{dusek}.
Clearly, the information is encoded in a way that does not depend on 
the value of $\phi_A$. In other words, this is a phase-reference 
independent way of encoding, and one could say that the BB84 protocol
makes use of {\em Qubits}.

On the other hand, one could imagine the 
state $|\alpha\exp(i\phi_A)\rangle$ shared in advance between Alice and Bob, 
and in that case this is akin to sharing a {\em refbit}. 
It is not quite the same,
as to lowest order in $|\alpha|$ the shared
state is $|0\rangle+\alpha\exp(i\phi_A)|1\rangle$, rather than an equal
superposition of $|0\rangle$ and $\exp(i\phi_A)|1\rangle$.

Similarly, the B92 protocol would have Alice send states to Bob 
of the form
\begin{equation}\label{qkd2}
|\beta \exp(i\phi_A)\rangle
|\alpha \exp(i(\phi_A+\phi_k))\rangle\,\,k=1,2,
\end{equation}
where $|\beta|\gg|\alpha|$ and there are only two phases chosen by Alice.
The idea is indeed to provide Bob with a full phase reference to allow him to 
unambiguously distinguish the two possible states with a reasonable 
(but not too large!)
probability. 
This method is more akin to sharing $N$ {\em refbits}, with $N$ large.
That method is wasteful in terms of resources (for other purposes
it would be better 
to use {\em Qubits} or
an encoding like that of Ref.~\cite{bartlett2}) but the large 
reference pulse is necessary for security purposes.
As an aside we note that creating superpositions of different
coherent states (even weak ones)
is very complicated in practice, so that  sending physical qubits
 this way is far from trivial.

For another practical example, we return (hopefully
for the last time) to the
discussions about teleportation with continuous variables (for details, 
see Refs.~\cite{telep}). In the language of the present paper, what
the typical teleportation experiment does  
is just to use many {\em refbits} to enable certain operations.
For example, Bob's unitary operation at the end of the teleportation
protocol must contain a phase $\phi_A$,
as it has been introduced by Alice's joint measurement.
In this same context, it was concluded in \cite{telep} 
that a certain mixture of two-mode squeezed states
in combination with a large phase reference pulse does possess distillable
entanglement, whereas the mixture 
without phase reference contains no entanglement. In the
language of our formalism, a two-mode squeezed state by itself contains
only {\em ebits}, but {\em refbits} in the form of
a laser acting as a phase reference can be used to generate
{\em Ebits}. Indeed,
the procedure used to distill the entanglement is essentially the same
as that used to
convert an {\em ebit}, 
in a single-shot protocol, to an {\em Ebit} by using a {\em refbit}.

Also, we have shown 
that (incoherent) teleportation, in the single-shot version, 
succeeds with
probability approaching unity only if one shares a large number of 
{\em refbits}. This then qualifies and quantifies the statement
in \cite{jmod} that in a teleportation protocol
one always needs to share a reference frame of some sort.

To return to more abstract concepts,
we have shown that 
the {\em cobit}, as introduced in \cite{harrow}, 
becomes a more powerful resource (relative to other resources) 
when no phase reference is shared. First of all, a {\em cobit}
 can always be used
to convert an unencoded resource into the encoded equivalent,
such as an {\em ebit} to an {\em Ebit}, or a {\em qubit} to a 
{\em Qubit}. Second,
with a reference frame present one has the equality
\begin{equation}
2\, cobits=1\, qubit+1\, ebit\,\,\,(*), 
\end{equation}
with the * indicating this equality holds when a phase reference is shared.
In contrast, without shared phase reference
we do have
\begin{equation}
2\,  cobits\geq1\, qubit+1\, ebit, 
\end{equation}
but only
\begin{equation}
1\,qubit+1\, ebit\geq 1\, Cobit.
\end{equation}
These relations show that the value of a {\em cobit} 
is exactly in between that of a {\em qubit} and an {\em ebit}
when there is a shared phase reference, but that it
moves closer to a {\em qubit}
the less of a shared phase reference one has.
The reason for a {\em cobit} to move closer
to a {\em qubit}
is that it can be implemented only by sending a physical qubit.
Being able to actually send a physical qubit becomes
more important in the absence of a reference frame, since one obviously
needs to send 
something in order to establish a reference frame.

Let us return to the incoherent version of superdense coding.
With $N$ large we found the relation
\begin{equation}\label{last}
1\, qubit+1\, ebit+N \, refbits\geq 
(2-1/\sqrt{\pi N})\, cbits.
\end{equation}
This follows from relation (\ref{super}) by substituting $p=1/2$
and replacing $N\rightarrow 2N$ and using $N$ refbits catalytically.
We note that
Alice and Bob could alternatively use $N$ {\em refbits} to estimate the 
phase difference $\phi_A-\phi_B$. 
Subsequently, Bob could then  
decode Alice's message by using his best estimate 
of the alignment of Alice's reference frame. That would transfer with 
high probability 2 classical
bits from Alice to Bob as well in a superdense coding protocol. 
The difference between 
the two approaches is that in one case they get sometimes only
1 classical
bit, but 
they know the bits are always correct; in the other case
 they always get 2 classical bits, 
but 1 bit might be incorrect and they do not know when.

Note, by the way, 
that using {\em refbits} to estimate the angle $\phi_A-\phi_B$
is the 2-D equivalent of the problem of estimating 
the Euler angles of a 3-D Cartesian
system by using spin-1/2 systems \cite{massar}.
But in the 2-D case, unlike in the 3-D case,
anti-parallel spins do not perform any better than
parallel spins, as there is an agreed-upon rotation axis
that can be used to convert parallel spins into anti-parallel spins.

Finally, we considered the 
difference between single-shot protocols and asymptotic
versions of the same protocols. In particular, encoding becomes a powerful
tool in the asymptotic limit. For instance, whereas one requires
an extra {\em cobit} in order 
to convert an {\em ebit} or a 
{\em qubit} into an encoded {\em Ebit} or {\em Qubit} in a single-shot
protocol, 
asymptotically one needs no extra resources to achieve the same conversion.
That is the same conclusion
 as reached in Ref.~\cite{bartlett2}, of course.
Note, though, that experiments in quantum communication typically do not 
implement the asymptotic version of protocols, but rather many
instances of single-shot protocols.

\end{document}